\documentclass[aps,preprint,showpacs,showkeys,amsmath,amssymb]{revtex4-1}
\usepackage{graphicx}
\usepackage{dcolumn}
\usepackage{bm}
\usepackage{hyperref}
\usepackage[T1]{fontenc}
\usepackage[utf8]{inputenc}

\begin{document}

\title{Quantum correlations of qubit-qutrit systems under general collective dephasing}
\author{Mazhar Ali}
\affiliation{Department of Electrical Engineering, Faculty of Engineering, Islamic University Madinah, 107 Madinah, Saudi Arabia}

\begin{abstract}
Most studies of collective dephasing for bipartite as well as multipartite quantum systems focus on a very specific orientation 
of magnetic field, that is, z-orientation. However, in practical situations, there are always small fluctuations in stochastic field and 
it is necessary that more general orientations of fields should be considered. We extend this problem to qubit-qutrit systems and study 
correlation dynamics for entanglement and local quantum uncertainty for some specific quantum states. 
We find that certain quantum states exhibit freezing dynamics both for entanglement and local quantum uncertainty. 
We analyze the asymptotic states and find the conditions for having non-zero entanglement and local quantum uncertainty. 
Our results are relevant for ion-trap experiments and can be verified with current experimental setups.  
\end{abstract}



\maketitle

\section{Introduction}
\label{Sec:intro}

Most realistic quantum systems are coupled to an environment that induces decoherence and dissipation. The study of such open quantum systems with 
different environments is an active area of research \cite{Aolita-review, Vega-RMP89-2017}. 
The correlations present in quantum states have potential applications, not limited to remote state preparation \cite{Dakic-Nat-2012}, 
entanglement distribution \cite{Streltsov-PRL108-2012, Chuan-PRL109-2012}, quantum metrology \cite{Modi-PRX1-2011}, quantum communication, and 
quantum computation \cite{Hayashi-Springer2015}. Considerable efforts have been devoted to develop a theoretical framework for quantification and 
characterization of quantum correlations \cite{gtreview, Chiara-RPP2018}. 
It is essential to analyze and simulate the effects of decoherence and dissipation on quantum correlations. In last two decades, detrimental 
effects of environment on quantum correlations have been studied extensively for bipartite as well as multipartite quantum systems
\cite{lifetime,Aolita-PRL100-2008,bipartitedec,Band-PRA72-2005,lowerbounds,Lastra-PRA75-2007, Guehne-PRA78-2008,Lopez-PRL101-2008, 
Ali-work, Weinstein-PRA85-2012,Ali-JPB-2014, Pramanic-PRA-2019}. 

To realize protocols of quantum information and quantum computation, a large variety of physical systems have been proposed and investigated. 
Electronic excitations of atoms and molecules, ion-traps, nuclear magnetic resonances, quantum dots, superconducting quantum interference 
devices, etc. \cite{Blatt-2006}. Of all these possibilities, ion-trap approach seems to be viable to realize quantum computer. 
In these experiments, ions/atoms are trapped in electromagnetic fields and quantum computations are performed by quantum logic gates, 
quantum measurements etc. \cite{Haeffner-PR469-2008}.
The typical noise here is caused by intensity fluctuations of electromagnetic fields which leads to collective dephasing process.  
Usually z-oriented magnetic field is considered to describe collective dephasing, such that Hamiltonian 
describing the interaction contain the corresponding dephasing operators. It is well known that decoherence degrades quantum correlations in general 
and entanglement in particular. The effects of collective dephasing on entanglement and other correlations have been investigated for bipartite and 
multipartite quantum systems \cite{Yu-CD-2002,AJ-JMO-2007,Li-EPJD-2007,Song-PRA80-2009,Ali-PRA81-2010,Karpat-PLA375-2011,Liu-PRA94-2016,Ali-EPJD-2017,
Karpat-CJP96-2018,Ali-MPLA-2019,Ali-BJP2019}. 

Recently, the specific $z$-oriented fluctuations in magnetic field has been extended to an arbitrary orientation for $N$ noninteracting 
atomic qubits \cite{Carnio-PRL-2015}. The resulting decoherence process can be called as {\it general collective dephasing}.   
It was shown that entanglement of a specific two qubits state may first decay up to some numerical value before suddenly stop decaying and 
maintain this stationary entanglement at all times \cite{Carnio-PRL-2015}. 
This non-trivial feature of entanglement decay is named as {\it freezing} dynamics of entanglement \cite{Ali-IJQI-2017}, where we found that 
specific quantum states of three and four qubits as well as most of the respective random states exhibit such dynamics under general collective 
dephasing. We note that there are no decoherence free subspaces (DFS) under general collective dephasing except $z$-oriented field. 
{\it Freezing} dynamics of entanglement has also shown to be present for $z$-oriented fields having DFS. 
Recently, it was found that under collective dephasing, some quantum states keep their entanglement locked. This 
peculiar phenomenon is termed as {\it time-invariant} entanglement. Indeed, some eigenvalues of quantum states may change at all times, 
however negative eigenvalues somehow are not functions of decay parameter, so entanglement is fixed for whole dynamics. Time-invariant 
entanglement was initially found for qubit-qutrit systems \cite{Karpat-PLA375-2011}. It has also been observed in experiments for qubit-qubit 
systems \cite{Liu-PRA94-2016}. We have studied this feature for multipartite systems and found no evidence of it for three qubits. 
Interestingly, we observed time-invariant entanglement for certain quantum states of four qubits \cite{Ali-EPJD-2017}. 
In another study, we investigated entanglement dynamics for qutrit-qutrit systems \cite{Ali-MPLA-2019} without finding any 
time-invariant entanglement but only freezing dynamics. 
For qubit-qutrit systems, we have shown that certain quantum states exhibit either time-invariant or sudden death 
of entanglement but never freezing dynamics. In addition, certain quantum states exhibit either freezing dynamics or sudden death but 
never time-invariant dynamics \cite{Ali-BJP2019}. In this work, we extend general collective dephasing to qubit-qutrit system such that 
qubit is exposed to stochastic fields in an arbitrary orientation $\vec{n}$, whereas qutrit is described by standard dephasing 
operator $\sigma_z$. We solve the system completely and provide a master equation in differential form which can be solved straight 
forwardly for any arbitrary initial quantum state.  
We study dynamics of entanglement and local quantum uncertainty for various orientations of magnetic field both for some specific quantum states. 
We also analyze quantum states at infinity and study quantum correlations of asymptotic states. We find the conditions that determine either finite 
end of entanglement or asymptotic entangled states. 
 
We organize this paper as follows. We present mathematical model in section~\ref{Sec:Model} and provide a compact master equation for 
qubit-qutrit system such that the general solution of it can be solved using any computer algebra system. We discuss quantification of  
entanglement and local quantum uncertainty in section~\ref{Sec:Ent}. In section~\ref{Sec:Res}, we study dynamics of these correlations
some specific quantum states and asymptotic states. We offer the summary in section~\ref{Sec:Cc}.

\section{General collective dephasing} 
\label{Sec:Model}

A qubit may be realized by a two level atom and a qutrit with a three level atom. 
This pair of atoms might share some quantum correlations caused by interaction in past and after that they are separated to ensure 
no coupling between them and treated as independent. However, they are allowed to interact with a noisy environment, collectively. 
The three level atom can be realized either in V-type configuration, or in $\Lambda$-type configuration, or in 
cascade configuration. The collective dephasing may appear due to coupling of atoms to stochastic magnetic field $B(t)$.
The Hamiltonian for this system can be considered as time dependent. As there are fluctuations in stochastic magnetic field $B(t)$, 
hence the ensemble average over it will lead to the decay parameter $\Gamma$.  
To our knowledge all previous studies on this problem usually take $z$-oriented magnetic field. The description of 
magnetic fields in an arbitrary direction is not studied before for qubit-qutrit systems. We fill this gap here and extend previous results  
by allowing an arbitrary orientation of magnetic field on qubit part. The Hamiltonian of the system (with $\hbar = 1$) can be written as  
\begin{eqnarray}
H(t) &=& - \frac{\mu}{2} \, \bigg[ \, B(t) \big(\vec{n} \cdot \vec{\sigma}^A + \sigma_z^B \big) \,  \bigg] \, , \nonumber \\ 
     &=& - \frac{\mu}{2} \, \bigg[ \, B(t) \big( n_x \, \sigma_x^A + n_y \, \sigma_y^A + n_z \, \sigma_z^A + \sigma_z^B \big) \,  \bigg] \, , \label{Ham} 
\end{eqnarray}
where $\mu$ is the gyro-magnetic ratio, $\vec{n} = n_x \, \hat{x} + n_y \, \hat{y} + n_z \, \hat{z}$, is a unit vector such that 
$|n_x|^2 + |n_y|^2 + |n_z|^2 = 1$, $\sigma_i^A$ are standard Pauli matrices for qubit A and $\sigma_z^B$ is dephasing operator for qutrit B. 
The stochastic function $B(t)$ denote statistically independent classical Markov processes and satisfy the following conditions:
\begin{eqnarray} 
\langle B(t) \, B(t')\rangle &=& \frac{\Gamma}{\mu^2} \, \delta(t-t') \,, \nonumber \\ 
\langle B(t)\rangle &=& 0 \, , \label{SC}
\end{eqnarray}
where $\langle \cdots \rangle$ is ensemble time average, $\delta(t-t')$ is delta function which vanished everywhere except $t = t'$, and $\Gamma$ 
denote the collective phase-damping rate.

The combined system-environment dynamics is given as
\begin{equation}
\rho_{st}(t) = U(t) \rho(0) U^\dagger(t)\,,
\end{equation}
where $\rho_{st}(t)$ is statistical density matrix for combined system. The unitary operator $U(t)$ is defined as
\begin{equation}
U(t) = \exp\bigg[-\mathrm{i} \int_0^t \, dt' \, H(t')\bigg]\,, 
\end{equation}
and $\rho(0)$ is the initial density matrix for combined system at $t = 0$. In this work, we do not allow any initial correlations between 
qubit-qutrit system and environment, such that they start with a product state, $\rho(0) = \rho_S \otimes \rho_R$, where $\rho_S$ is density 
matrix for qubit-qutrit system and $\rho_R$ is density matrix of environment.   
The density matrix for qubit-qutrit system alone can be calculated by first taking ensemble average and then partial trace over the noisy field. 
The time evolved density matrix of qubit-qutrit system can be obtained in several ways. We prefer the master equation approach. In our 
recent work \cite{Ali-BJP2019}, we have provided the detailed mathematical derivation on how to get a master equation. We follow the same procedure 
and after considerable simplification, we get the master equation, given as
\begin{eqnarray}
\dot{\rho}(t) &=& - \frac{\Gamma}{4} \, \bigg( \sum_{i,j = x,y,z} \, n_i \, n_j \, \big( \sigma_i^A \, \sigma_j^A \, \rho(t) 
+ \rho(t) \, \sigma_i^A \, \sigma_j^A - 2 \, \sigma_i^A \, \rho(t) \, \sigma_j^A \, \big) \nonumber \\&& - 2 \, \sum_{j = x,y,z} \, n_j \, 
\big( \sigma_j^A \, \rho(t) \, \sigma_z^B + \sigma_z^B \, \rho(t) \, \sigma_j^A  
- \frac{1}{2} \, \big\{ \, \{\sigma_j^A , \sigma_z^B\}, \rho(t) \, \big\} \, \big) \, \nonumber \\&& 
+ \, \{ \, \sigma_z^B \, \sigma_z^B \,,  \rho(t) \, \} - 2 \, \sigma_z^B \, \rho(t) \, \sigma_z^B \, \bigg)\,, \label{DME} 
\end{eqnarray}
where $\{ \hat{g}, \hat{h} \} = \hat{g} \, \hat{h} + \hat{h} \, \hat{g}$ is anti-commutator defined for two operators. 
This is a simple differential equation, which can be exactly solved using any computer algebra system. 
We have obtained the general solution which is quite cumbersome to present here. Nevertheless, we just mention that there are 
groups of four matrix elements coupled together, like $(\rho_{11}(t)$, $\rho_{14}(t)$, $\rho_{41}(t)$, $\rho_{44}(t))$, another 
$(\rho_{22}(t)$, $\rho_{25}(t)$, $\rho_{52}(t)$, $\rho_{55}(t))$ etc. 
For general collective dephasing $(n_x \neq 0$, $n_y \neq 0$, and $n_z \neq 0)$, there are no decoherence free subspaces (DFS) \cite{Yu-CD-2002} 
in this system. Another interesting property of the dynamics is the fact that all initially zero matrix elements may not remain zero at later times.
We note that for $z$-orientation of magnetic field, that is, $n_x = n_y = 0$, and $n_z = 1$, we recover all our earlier results \cite{Ali-BJP2019}. 

In calculating above equation, we have defined the computational basis $\{ \, |0,0\rangle$, $|0,1\rangle$, $|0,2\rangle$, $|1,0\rangle$, 
$|1,1\rangle$, $|1,2\rangle \, \}$, where the first basis is for qubit and second basis is for qutrit. 
We have also ignored the subscripts A and B and used the notation $|0 \rangle \otimes |0\rangle = |0 \, 0 \rangle$ throughout our work. 

\section{Entanglement and quantum local uncertainty}
\label{Sec:Ent}

The problem of separability for Hilbert spaces with dimension $4$ (qubit-qubit) and $6$ (qubit-qutrit) has been solved. 
Peres's criterion states that for separable states the partial transpose with respect of any one of the subsystem has all positive eigenvalues 
\cite{Peres-PRL-1996}. A quantum state is entangled if it has at least one negative eigenvalue of its partially transposed matrix. 
This condition is necessary and sufficient for $2 \otimes 2$ and $2 \otimes 3$ systems. For higher  
dimensional Hilbert spaces, there may exist entangled states having positive partial transpose \cite{Horodecki-RMP-2009}. 
This criterion lead to a measure of entanglement, negativity, defined as the sum of absolute values of all possible negative 
eigenvalues \cite{Vidal-PRA65-2002}. For a given quantum state $\rho$, negativity is given as  
\begin{equation}
N (\rho) = 2 \, \bigg( \, \sum_i \, | \zeta_i| \, \bigg)\, ,
\end{equation}
where $\zeta_i$ are negative eigenvalues. The factor $2$ is for normalization. There are several other measures of 
entanglement \cite{Horodecki-RMP-2009} but it is not known how to compute them except for some special cases. 
For qubit-qutrit system, it may appear easy to compute negativity, however, in most cases analytical expressions are not possible. 
The reason is presence of large number of parameters in density matrix. For example, 
in our current work we have at least two real parameters related with orientation of magnetic field ($n_i$), minimum of one or two parameters with 
initial quantum states and a real parameter $\Gamma t$ related with decoherence. To find analytical eigenvalues 
of a $6 \times 6$ matrix with $4$ or larger parameters is a hard task. Nevertheless, the solution to this 
problem exists in literature. Recently, a powerful technique has been worked out to detect and characterize entanglement \cite{Bastian-PRL106-2011}. 
The method is to use positive partial transpose mixtures (PPT mixtures), for details, see \cite{Bastian-PRL106-2011}. 

Another quantum correlation which we study here is recently defined measure, called local quantum uncertainty (LQU). This measure 
has a compact formula for $2 \otimes d$ systems \cite{Girolami-PRL110-2013}. This measure is similar to quantum discord. 
To calculate this measure, we need to find the maximum eigenvalue of $3 \times 3$ matrix. This is simple job for less number of parameters in a 
matrix. LQU is defined as the minimum skew information obtained via local measurement on qubit, that is,  
\begin{equation}
\mathcal{Q}(\rho) \equiv \, \min_{M_A} \, \mathcal{J} (\rho , M_A \otimes \mathbb{I}_B) \,, 
\end{equation}
where $M_A$ is self-adjoint matrix on subsystem $A$, and $\mathcal{J}$ is the skew information \cite{Luo-PRL91-2003} defined as 
\begin{equation}
\mathcal{J} (\rho , \, M_A \otimes \mathbb{I}_B) \, = \,- \frac{1}{2} \, \rm{Tr} ( \, [ \, \sqrt{\rho}, \, M_A \otimes \mathbb{I}_B ]^2 \, ) \,.
\end{equation}
For qubit-qudit systems, local quantum uncertainty is given as
\begin{equation}
\mathcal{Q}(\rho) = 1 - \, \rm{max} \, \{ \delta_1 \,, \delta_2 \, , \delta_3 \, \}\,, 
\end{equation}
where $\delta_i$ are eigenvalues of $3 \times 3$ matrix $\mathcal{R}$, with elements defined as 
\begin{equation}
 r_{ij} \equiv \rm{Tr} \, \big\{ \, \sqrt{\rho} \, (\sigma_i \otimes \mathbb{I}_B) \, \sqrt{\rho} \, (\sigma_j \otimes \mathbb{I}_B) \, \big\}\,, 
\end{equation}
where $\sigma_i$ are Pauli matrices. We have extended this measure for multi-qubits systems as well \cite{Ali-EPJD-2020}. 

\section{Correlation dynamics of specific states}
\label{Sec:Res}

Being equipped with general solution for any arbitrary initial state and methods to compute entanglement and LQU, we now proceed to 
study how quantum states react to an arbitrary magnetic field as compared with a special z-orientation. To this aim, we choose some 
specific quantum states, well known in literature and study their behavior for various settings of $\vec{n}$. 
First we take quantum states with two real parameters $\alpha$ and $\gamma$ defined for $2 \otimes d$ quantum 
systems \cite{Chi-JPA36-2003}. It is known that an arbitrary quantum state $\rho$ in $2 \otimes d$ can be transformed 
to $\rho_{\alpha, \gamma}$ using local operations and classical communication (LOCC). For our system, the states are given as 
\begin{eqnarray}
\rho_{\alpha, \gamma} &=& \alpha \, (| 0\, 2 \rangle\langle 0 \, 2| + | 1\, 2 \rangle\langle 1 \, 2|) + \beta \, ( | \phi^+ \rangle\langle \phi^+| 
+ | \phi^- \rangle\langle \phi^-|  
\nonumber \\&& + | \psi^+ \rangle\langle \psi^+| \, ) + \gamma \, | \psi^- \rangle\langle \psi^-| \,,
\label{Eq:rhoag}
\end{eqnarray}
where 
\begin{eqnarray}
| \, \phi^\pm \rangle &=& \frac{1}{\sqrt{2}} \, (\, |0\, 0\rangle \pm |1\,1 \rangle \, ) \\
| \, \psi^\pm \rangle &=& \frac{1}{\sqrt{2}} \, (\, |0\, 1\rangle \pm |1\,0 \rangle \, ) \, , 
\end{eqnarray}
and parameter $\beta$ is dependent on $\alpha$ and $\gamma$ by the unit trace condition, 
\begin{eqnarray}
 2 \, \alpha + 3 \, \beta  + \gamma = 1 \, .
\end{eqnarray}
From Eq.~(\ref{Eq:rhoag}) one can easily obtain the range of parameters as  $0 \leq \alpha \leq 1/2$, $ 0 \leq \beta \leq 1/3$, and 
$0 \leq \gamma \leq 1$. We note that the states of the form $\rho_{0, \gamma}$ are equivalent to Werner states \cite{Wer-PRA89} in a 
$2 \otimes 2$ quantum systems. Moreover, the states $\rho_{\alpha, \gamma}$ have the property that their PPT (positive partial transpose) 
region is always separable \cite{Chi-JPA36-2003}. 

\begin{figure}[t!]
\scalebox{2.30}{\includegraphics[width=1.99in]{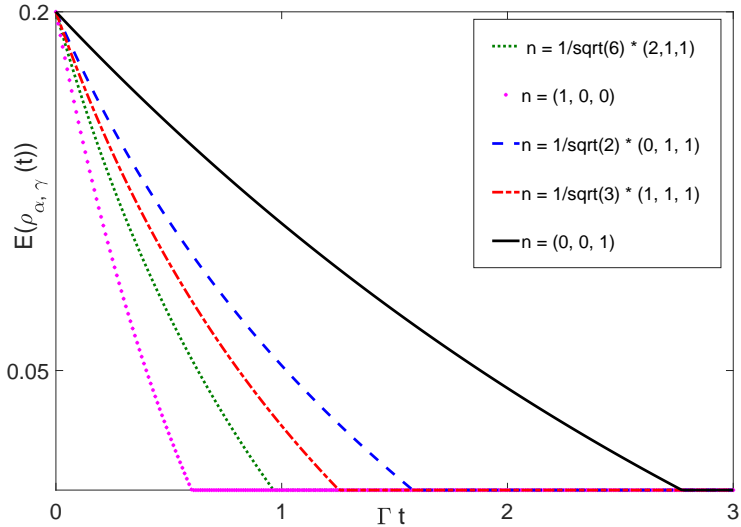}}
\caption{Negativity $E (\rho_{\alpha, \gamma}(t))$ is plotted against parameter $\Gamma t$. It can be seen that for all settings of magnetic field, 
entanglement is lost at finite times. See text for details.}
\label{Fig:1}
\end{figure}
Figure~(\ref{Fig:1}) depicts dynamics of entanglement for state $\rho_{\alpha, \gamma}(t)$ with five settings of $\vec{n}$. 
We have taken $\alpha = 0.1$, $\beta = 0.1$, and $\gamma = 0.5$. The solid (black) line is 
for z-orientation of magnetic field. We observe that for this specific quantum state, entanglement is lost at finite times for all settings of 
$\vec{n}$. The general orientations of magnetic field have severe effect on entanglement as it is lost at earlier times. 
These states are quite fragile under general collective dephasing.  

\begin{figure}[t!]
\scalebox{2.30}{\includegraphics[width=1.99in]{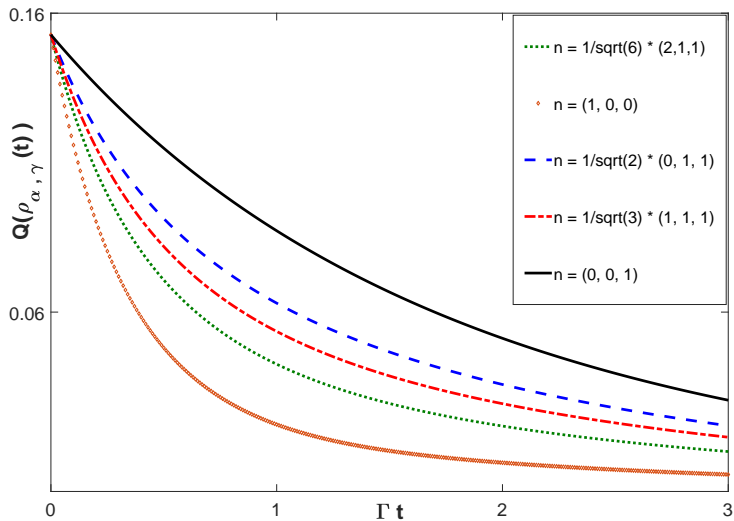}}
\caption{LQU $Q (\rho_{\alpha, \gamma}(t))$ is plotted against parameter $\Gamma t$. It can be seen that for all settings of magnetic field, 
LQU tends to freeze. See text for details.}
\label{Fig:2}
\end{figure}
In Figure~(\ref{Fig:2}), we show LQU against decay parameter $\Gamma t$ for same settings of $\vec{n}$ as in Figure~(\ref{Fig:1}). There is a common 
behavior with these two correlations that the order of decay is similar. The slowest decay is for z-orientation and faster decay for other orientations. 
However, LQU tends to stabalize to a specific value. This behaviour is similar to freezing dynamics of correlations. We will explain the reason for these
kind of trajectories later in this work. 

Let us consider another example. We first define a single parameter class of states as
\begin{equation}
\tilde{\rho}_\alpha = \alpha \, |\psi_1 \rangle\langle \psi_1| + \frac{1 - \alpha}{6} \, \mathbb{I}_6 \,, 
\end{equation}
where $\mathbb{I}_6$ is $6 \times 6$ identity matrix and  $0 \leq \alpha \leq 1$. In this equation, the pure state $| \psi_1\rangle$ is another 
maximally entangled state as  
\begin{equation}
|\psi_1 \rangle = \frac{1}{\sqrt{2}} \, (|0 \, 0\rangle + |1 \, 2 \rangle)\, .
\end{equation}
Such states are called isotropic states and they are NPT for $1/4 < \alpha \leq 1$, and hence entangled.
We can now define a two parameter family of states, which are mixture of isotropic states and $|\psi_3\rangle$, given as 
\begin{equation}
\rho_{\alpha,\beta} = \beta \, |\psi_3\rangle \langle \psi_3 | + (1 - \beta) \, \tilde{\rho}_\alpha \, ,    
\end{equation}
where  $0 \leq \beta \leq 1$. In Ref.\cite{Ali-BJP2019}, we have observed time-invariant entanglement for this state. Here we want to study how these 
states change their correlations under general collective dephasing. 

\begin{figure}[t!]
\scalebox{2.30}{\includegraphics[width=1.99in]{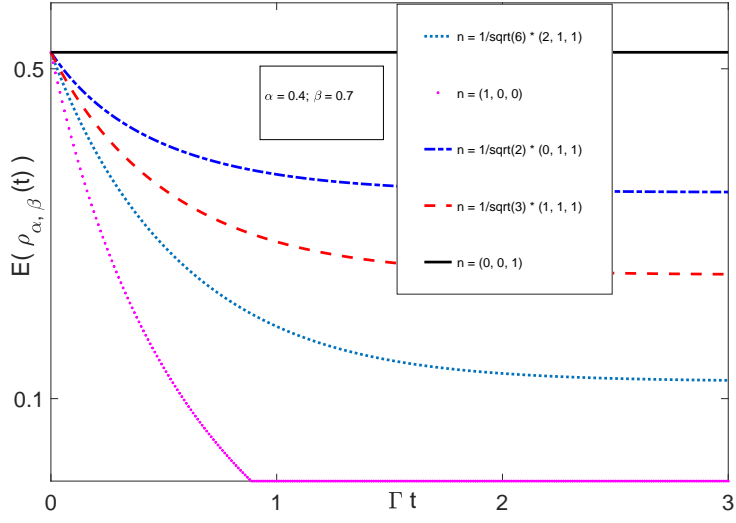}}
\caption{Negativity for an initial state $\rho_{\alpha, \beta}(t)$ is plotted against decay parameter $\Gamma t$ for various values of  
$\vec{n}$. We have taken $\alpha = 0.4$ and $\beta = 0.7$. It can be seen that in one choice we get time-invariant entanglement, in another choice, 
we get sudden death of entanglement and in three other choices, we get freezing dynamics of entanglement.}
\label{Fig:3}
\end{figure}
In Figure~(\ref{Fig:3}), we plot negativity against decay parameter $\Gamma t$  for same choices of $\vec{n}$. We have taken 
$\alpha = 0.4$ and $\beta = 0.7$. For $n_x = n_y = 0$, and $n_z = 1$, solid (black) we see time-invariant entanglement 
as we have observed earlier \cite{Ali-BJP2019}. For $n_y = n_z = 0$, and $n_x = 1$ starred (pink) line, we observe sudden death of entanglement 
at $\Gamma t \approx 0.9$. This specific orientation of magnetic field seems to be most destructive for 
entanglement. Another interesting feature is the appearance of freezing dynamics of entanglement for other three 
settings of $\vec{n}$. We can see that dashed-dotted (blue) line, dashed (red) line, and dotted line lead to freezing dynamics of entanglement. 
We will analyze asymptotic quantum states below and based on the eigenvalues of the partially transposed matrix, 
we will explain this behavior of entanglement dynamics for these states. 

\begin{figure}[t!]
\scalebox{2.30}{\includegraphics[width=1.99in]{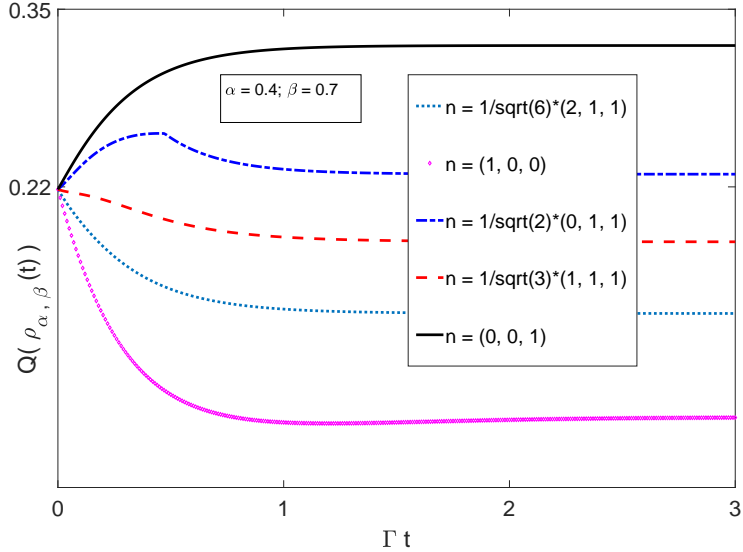}}
\caption{LQU $Q (\rho_{\alpha, \gamma}(t))$ is plotted against parameter $\Gamma t$. It can be seen that for $n_x = 0$, LQU first increase before 
exhibiting freezing dynamics. See text for details.}
\label{Fig:4}
\end{figure}
Figure~(\ref{Fig:4}) shows LQU against parameter $\Gamma t$ for same settings of $\vec{n}$. Here, we observe quite interesting dynamics of local 
quantum uncertainty as in two settings, LQU first increase and then exhibit freezing dynamics. Whereas for other three settings of $\vec{n}$, we observe 
slight decay before showing freezing dynamics of LQU. The increment in LQU is for both settings which have $n_x = 0$. We have analyzed the 
matrix elements of general solution and also at infinity and have found that that for this choice of $\vec{n}$, there are some matrix elements which 
are initially zero but become non-zero during dynamics. It is due to sudden appearance of these "coherences" that LQU increases and later show 
freezing dynamics. 

The most general solution suggests that it is possible to analyze asymptotic quantum states. By taking $\Gamma t \to \infty$, 
we obtained quantum states at infinity as 
\begin{eqnarray}
\rho(\infty) = \left(
\begin{array}{llllll}
\varrho_{11} & 0 & \varrho_{13} & \varrho_{14} & 0 & \varrho_{16} \\
0 & \varrho_{22} & 0 & 0 & \varrho_{25} & 0 \\
\varrho_{31} & 0 & \varrho_{33} & \varrho_{34} & 0 & \varrho_{36} \\
\varrho_{41} & 0 & \varrho_{43} & \varrho_{44} & 0 & \varrho_{46} \\
0 & \varrho_{52} & 0 & 0 & \varrho_{55} & 0 \\
\varrho_{61} & 0 & \varrho_{63} & \varrho_{64} & 0 & \varrho_{66}
\end{array}
\right) \,,\label{Eq:INF}
\end{eqnarray}
where $\varrho_{ij}$ are function of parameters $n_k$ and initial density matrix elements $\rho_{qr}$. We noticed that for special case of 
$z$-orientation, that is, $n_x = n_y = 0$, and $n_z = 1$, we have all off-diagonal elements equal to zero except 
$\varrho_{34}$ and $\varrho_{43}$ (presence of DFS). For two other special orientations, ($n_x = 1$, $n_y = n_z = 0$) and 
($n_y = 1$, $n_x = n_z = 0$) and in general for $n_x \neq n_y \neq n_z \neq 0$, we have all $\varrho_{ij}$ as non-zeros. Hence there are no 
DFS for general collective dephasing. This structure already suggest the possibility of entangled states at infinity. 

We can identify types of entangled states which must exhibit finite end of entanglement. For other types of entangled
states, we may get either freezing dynamics or abrupt end. Consider the various entangled pure states in Schmidt 
decomposition 
\begin{equation}
| \Phi_1 \rangle = \alpha_1 \, | 0\,0\rangle \pm \beta_1 \, | 1 \, 2\rangle\,,
\label{Eq:Ph1}
\end{equation}
\begin{equation}
| \Phi_2 \rangle = \alpha_2 \, | 0\,0\rangle \pm \beta_2 \, | 1 \, 1\rangle\,,
\label{Eq:Ph2}
\end{equation}
\begin{equation}
| \Phi_3 \rangle = \alpha_3 \, | 0\,1\rangle \pm \beta_3 \, | 1 \, 2\rangle\,,
\label{Eq:Ph3}
\end{equation}
\begin{equation}
| \Phi_4 \rangle = \alpha_4 \, | 0\,1\rangle \pm \beta_4 \, | 1 \, 0\rangle\,,
\label{Eq:Ph4}
\end{equation}
\begin{equation}
| \Phi_5 \rangle = \alpha_5 \, | 0\,2\rangle \pm \beta_5 \, | 1 \, 0\rangle\,,
\label{Eq:Ph5}
\end{equation}
\begin{equation}
| \Phi_6 \rangle = \alpha_6 \, | 0\,2\rangle \pm \beta_6 \, | 1 \, 1\rangle\,.
\label{Eq:Ph6}
\end{equation}
It should be mentioned here that an arbitrary pure state for qubit-qutrit system can be written as  
\begin{equation}
|\Phi \rangle = (U_A \otimes U_B) \, \big( \alpha \, |0\,0 \rangle + \sqrt{1 - \alpha^2} \, | 1 \, 1\rangle\,\big)\,, 
\end{equation}
where $U_A$ and $U_B$ denote transformations from the computational basis to the Schmidt basis on qubit and qutrit, respectively. A close 
examination of asymptotic states suggest that any mixed quantum state which has entangled states $|\Phi_2\rangle$ (Eq.~\ref{Eq:Ph2}), 
$|\Phi_3\rangle$ (Eq.~\ref{Eq:Ph3}), $|\Phi_4\rangle$ (Eq.~\ref{Eq:Ph4}), and $|\Phi_6\rangle$ (Eq.~\ref{Eq:Ph6}) as a dominant fraction 
in it, necessarily exhibit entanglement sudden death. Whereas mixed states with large fractions of states $|\Phi_1\rangle$ (Eq.~\ref{Eq:Ph1}) and 
$|\Phi_5\rangle$ (Eq.~\ref{Eq:Ph5}) can lead to either sudden death of entanglement or freezing dynamics of entanglement. We also note that 
$|\Phi_2\rangle$ and $|\Phi_4\rangle$ are related with each other by a local switch on qutrit alone. Similar relation is also between 
$|\Phi_3\rangle$ and $|\Phi_6\rangle$ and also between $|\Phi_1\rangle$ and $|\Phi_5\rangle$. We demonstrate below the possibility of entangled 
states at infinity under general collective dephasing. 

Our first example above has $|\Phi_2\rangle$ (Eq.~\ref{Eq:Ph2}) and $|\Phi_4\rangle$ (Eq.~\ref{Eq:Ph4}) 
in it with $\alpha_i = \beta_i = 1/\sqrt{2}$, so we must have sudden death of entanglement for all orientations of magnetic field. 
This is precisely what we have observed in Figure~(\ref{Fig:1}). 
Our second example is mixture of $|\Phi_1\rangle$ with $\alpha_1 = \beta_1 = 1/\sqrt{2}$ and $|\Phi_5\rangle$ with 
$\alpha_5 = \beta_5 = 1/\sqrt{2}$. Therefore, we may get either sudden death of entanglement or freezing dynamics depending upon orientation 
of magnetic field. We note that time-invariant entanglement can only occur for special orientation of $\vec{n} = (0,0,1)$. 
It is simple to check the possible negative eigenvalues for the respective asymptotic state and discuss the dynamical behavior.

\section{Conclusions}
\label{Sec:Cc}

Qubit-qutrit systems have been studied in a variety of quantum systems \cite{Metwally-QIC-2016}. 
We have extended the previous studies on collective dephasing of qubit-qutrit systems from a specific orientation of magnetic field to 
an arbitrary orientation. The usual $z$-oriented magnetic field give rise to decoherence free subspaces (DFS), whereas 
general collective dephasing have no such spaces. We have considered our qubit interacting with stochastic fields oriented in an arbitrary direction 
dictated by $\vec{n}$. We have obtained a master equation and its solution for an arbitrary 
initial quantum state. As we studied recently, entanglement in general has three non-trivial types of dynamics, namely, time-invariant entanglement, 
sudden death of entanglement, and freezing dynamics of entanglement. It has been observed that for qubit-qubit systems interacting with general 
directions of magnetic field \cite{Carnio-PRL-2015}, one can find freezing dynamics of entanglement instead of specific $z$-direction where we can 
also find time-invariant feature. In this work, we have seen that for specific quantum states, general collective dephasing degrades entanglement 
more than specific $z$-oriented field. Even for quantum state exhibiting time-invariant entanglement under specific $z$-oriented magnetic field, we 
get either sudden death of entanglement or freezing dynamics of entanglement. 
We have also studied local quantum uncertainty for various settings of magnetic field $\vec{n}$. We have found that 
these states exhibit freezing dynamics of LQU for all orientations of field. For certain quantum states, LQU can increase instead of decaying and later 
show freezing dynamics only for $n_x = 0$. We found that the reason for this behavior is sudden appearance of matrix elements which were initially zero. 
Such coherences can increase the amount of LQU. However, these matrix elements do not effect entanglement. 
We were also able to find most general asymptotic quantum states in terms of parameters $n_i$ and initial density matrix elements. This knowledge can 
conclusively explain the dynamics of entanglement and LQU for any arbitrary orientation of magnetic field. As many experiments are already in quite 
advanced stage for ion-traps, where this kind of noise is dominant, we believe that our study is relevant to such experiments and these observations 
can be readily demonstrated. Future avenues could be to develop general collective dephasing for high dimensional quantum systems.  

\section*{References}

\end{document}